\documentclass{article}
\usepackage{epsf}

\def\beq{\begin{equation}}
\def\eeq{\end{equation}}
\def\bea{\begin{eqnarray}}
\def\eea{\end{eqnarray}}

\relax
\begin{document}

\begin{titlepage}

\vskip 2cm

\begin{center}

{\Large\bf Supersymmetric Model of Spin-1/2 Fermions
  on a Chain}\\[.5in]

{\large\bf  Raoul Santachiara and Kareljan Schoutens}\\[1mm]

{\large\it  Institute for Theoretical Physics, Valckenierstraat 65 \\[.2mm]
            1018 XE  Amsterdam, THE NETHERLANDS}\\[1cm]

March 14, 2005 \hfill \\[2cm]

{\large\sc Abstract}

\end{center}

\begin{quote}
In recent work, $N=2$ supersymmetry has been proposed as a tool
for the analysis of itinerant, correlated fermions on a 
lattice. In this paper we extend these considerations to the
case of lattice fermions with spin 1/2 . We introduce a model for 
correlated spin-1/2 fermions with a manifest $N=4$ supersymmetry, 
and analyze its properties. The supersymmetric ground states that 
we find represent holes in an anti-ferromagnetic background.
\end{quote}

\end{titlepage}

\newpage

\section{Introduction}

For many condensed matter systems, the key to understanding the 
physical properties lies in the analysis of a quantum many body
problem with strong correlations. For the analysis of such systems,
approaches that go beyond the standard perturbative techniques are 
always needed. It has recently been proposed 
\cite{Fendley:2002sg,Fendley:2003je} 
that, for a special class of lattice models for correlated fermions, 
supersymmetry can provide a tool for non-perturbative analysis. In these
models, questions about the existence and degeneracies of strongly 
correlated ground states at zero energy are easily answered with the 
help of supersymmetry and elementary combinatorics. Explicit properties
of these same ground states are being studied with techniques that
are, in various ways, associated with supersymmetry 
\cite{Fendley:2002sg,Pearce,Beccaria}.

In the formulation of supersymmetric lattice models, the starting point is 
the definition of two nilpotent
fermionic operators denoted $\overline{Q}$ and $Q=\overline{Q}^{\dag}$. The
Hamiltonian $H$ is then built from these two hermitian conjugate
supercharges as  $H=\{\overline{Q},Q\}$ (this is often called $N=2$
supersymmetry as there are two supersymmetry generators). This special 
algebraic structure implies a number of important properties, which lead 
to considerable computational simplification. In particular, the ground
state structure of these of $H$ can be analyzed with help of
relatively simple combinatorics and of cohomology theory. It is important 
to emphasize that this approach remains well valid also in dimension $D>1$ 
\cite{Fendley:2005,Eerten} where, in general, very few non-perturbative 
techniques are
available. The study of two-dimensional supersymmetric models in dimension
$D>1$ has revealed the existence of a large number of quantum systems 
characterized by a finite ground state entropy at zero temperature, 
providing an intriguing new realization of the phenomenon of quantum 
frustration \cite{Fendley:2005} .

Further results have been obtained by studying a family of one-dimensional 
supersymmetric models, denoted with $M_k[\{x_a\}]$, which depend on $k-1$ 
free parameters $x_a$. These models $M_k$ 
are described in terms of spinless fermions on a chain. The Hilbert space 
is restricted so that no more than $k$ consecutive sites can be occupied.  
It was found \cite{Fendley:2003je} that there is a close connection 
between these models and a number of models that are well-known
in condensed matter theory: $M_1$ is closely related to the $XXZ$ chain at 
$\Delta=-1/2$, $M_2[x=0]$ connects to the $su(2|1)$-supersymmetric $t-J$ 
model, and the general model $M_k[\{x_a\}]$ can be related to a spin-$k/2$ 
$XXZ$ model. 
These relations have permitted to argue that, at some particular coupling 
$\{x_a\}$, the model $M_k$ is described in the continuum limit by the 
$k$-th minimal model of $N=2$ superconformal field theory. With this result, 
all minimal universality classes of critical behavior with $N=2$
superconformal symmetry have been represented by lattice models with
explicit $N=2$ supersymmetry on a discrete, finite lattice.

Inspired by these results, and by the obvious wish to make a connection
with lattice models for strongly correlated electrons, we have investigated 
possible generalizations to spin-1/2 fermions. Insisting on 
$SU(2)$ spin symmetry, one is quickly led to an algebraic structure
having $N=4$ rather than $N=2$ supersymmetries. In this paper we
follow this line of thought, and present an $N=4$ supersymmetric lattice 
model for itinerant spin-1/2 fermions in one spatial dimension. 
For small lattices we find explicit supersymmetric ground states, showing
that supersymmetry is unbroken. 

Our construction is rather involved, and in its present form it is
restricted to one spatial dimension. Nevertheless, our results do 
illustrate the potential use of $N=4$ supersymmetry for the analysis of 
antiferromagnets that are doped with holes. As such they invite further 
analysis, in particular in the direction of models in $D=2$ or higher 
spatial dimensions, where for $N=2$ supersymmetric lattice
models remarkable results have been obtained \cite{Fendley:2005}.

\section{Lattice model with extended supersymmetry}

In this main section we present in a number of steps the construction
of our $N=4$ supersymmetric lattice model for spin-1/2 fermions.
We begin by specifying how the spin $SU(2)$ algebra and supersymmetry
give rise to $N=4$ extended supersymmetry (section 2.1). We then
specify a constrained Hilbert space for spin-1/2 fermions
(section 2.2) and define the action of the four supercharges
in section 2.3. Finally, in section 2.4, we present the Hamiltonian
and the supersymmetric ground states that we obtained.

\subsection{Algebraic structure}

The first step is to fix the  algebra formed by the supersymmetry and
the $SU(2)$ generators. There are two couples of hermitian conjugate 
supercharges,  $\overline{Q}_{\alpha}$ and $Q_{\alpha}$, that add or take out 
a fermion in the spin state $\alpha$, with $\alpha=\uparrow, \downarrow$.  
These operators are nilpotent, {\it i.e.}\ they obey 
$(\overline{Q}_{\alpha})^2=(Q_{\alpha})^2=0$.  The generators of the spin 
symmetry, which we denote by $J^{a}$ ($a=+,-,0$), can appear in the 
anti-commutation relations of the supercharges. In its simplest form,
the so called $SU(2)$-extended or $N=4$ supersymmetry algebra reads
\bea
&& \{\overline{Q}_{\alpha},\overline{Q}_{\beta}\}=\{Q_{\alpha},Q_{\beta}\}=0 \ ,
\label{comm1}
 \\
&& \{\overline{Q}_{\alpha},Q_{\beta}\} = \delta_{\alpha,\beta} H
 +\gamma \, \sigma^{a}_{\alpha,\beta} J^{a}, \quad
 \alpha,\beta=\uparrow,\downarrow \ , \quad a=+,-,0 \ ,
\label{important_relation}
\eea
where the $\sigma^{a}_{\alpha,\beta}$ are the Pauli matrices, $\gamma$
is a constant and $H$ will be identified as the Hamiltonian 
of the supersymmetric theory.
One can verify, by using the Jacobi identities, that the above relations
imply the following commutation relations
\beq
[J^{a},\overline{Q}_{\alpha}]=\sigma^{a}_{\alpha,\beta} \overline{Q}_{\beta} \ ,
\qquad 
\left[J^{a},J^{b}\right]= f^{abc} J^{c} \ ,
\eeq
where $f^{abc}$ are the structure constants of the $SU(2)$
algebra which is thus included as a sub-algebra.

\begin{figure}
\centerline{
\epsfxsize=10.0cm
\epsfclipon
\epsfbox{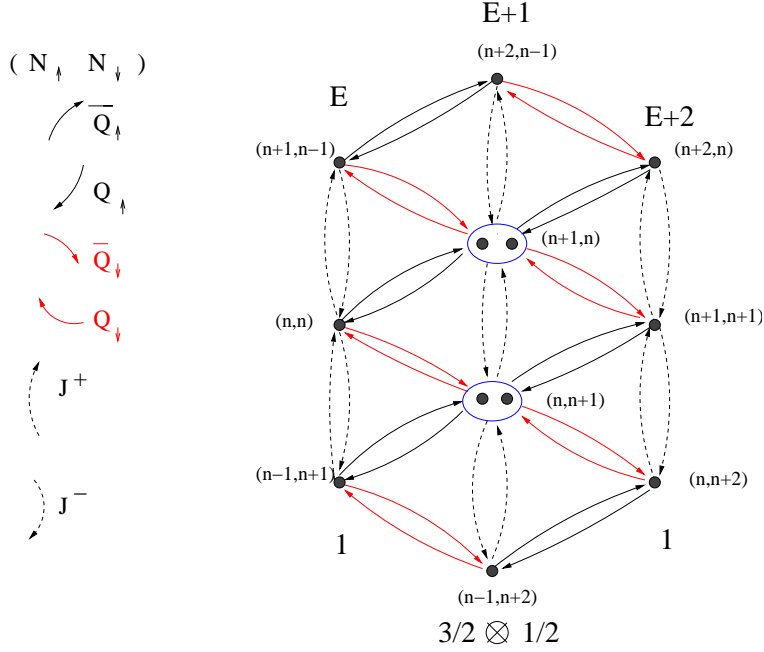}}
\caption{
A characteristic supermultiplet forming a representation of the
algebra (\ref{comm1})-(\ref{important_relation}) for $\gamma=1$. 
Each state (dot) in the supermultiplet is characterized by a pair 
$(N_{\uparrow},N_{\downarrow})$ which indicates the number of spin 
$\uparrow$ and $\downarrow$ fermions.
The arrows show the action of the supersymmetry and $SU(2)$ generators on 
such states. This particular supermultiplet includes $SU(2)$ multiplets of 
spin $1/2$, 1, 1 and $3/2$.}
\label{multiplet}
\end{figure}

The energy operator
\beq
H=\{\overline{Q}_{\uparrow},Q_{\uparrow}\}+
              \{\overline{Q}_{\downarrow},Q_{\downarrow}\}  
\label{hamiltonian}
\eeq
defined in (\ref{important_relation}) satisfies the relations
\beq
\left[H,J^{a}\right]=0 \ ,
\qquad
\left[H,\overline{Q}_{\alpha}\right]=\gamma \, \overline{Q}_{\alpha} \ ,
\qquad 
\left[H,Q_{\alpha}\right]=-\gamma \, Q_{\alpha} \ .
\eeq
From these relations, it follows that $H^{'}=
H- {\it \hat{N}}$, where ${\it \hat{N}}$ is the total 
fermion-number operator, commutes with all generators of the $N=4$ 
supersymmetry algebra. This implies that $H^{'}$
will be constant on supermultiplets built from the generators 
of the algebra. A characteristic supermultiplet is shown in Fig.~\ref{multiplet}.
It is composed of SU(2)-spin multiplets of spin $1/2$, 1, 1
and $3/2$. The hamiltonian $H$ is invariant 
under $SU(2)$-spin and it changes by $\pm \gamma$ under the action
of one of the supercharges.

By fixing the normalization of the supercharges, we have three
different algebras (\ref{important_relation}), which correspond to the
cases $\gamma =0$, $\gamma=-1$ and $\gamma=1$.

When  $\gamma=0$, the eqs.~(\ref{comm1})-(\ref{important_relation})
correspond to the usual
Clifford algebra which is trivially satisfied by taking
$\overline{Q}_{\alpha}=\sum_i c^{+}_{i,\alpha}$, where  $c^{+}_{i,\alpha}$
creates a fermion at site $i$ with spin $\alpha$. The corresponding 
Hamiltonian is simply a constant, $H=L$, with $L$ the
number of lattice sites. We have not found any other definition of the 
supercharges which respect eqs.~(\ref{comm1})-(\ref{important_relation}) 
for $\gamma=0$.

If we restrict the Hilbert space by allowing at most one fermion at
each site, we can verify that the supercharges
$\overline{Q}_{\alpha}=\sum_i c^{+}_{i,\alpha}(1-n_i)$, with $n_i$ the
fermion number operator at site $i$, and their
conjugates form  the algebra (\ref{comm1})-(\ref{important_relation})
with $\gamma=-1$. The associated Hamiltonian is 
$H=2L-{\it \hat{N}}$  and it is trivially 
diagonalizable.

Quite surprisingly, we have found a non-trivial realization of the 
algebra (\ref{comm1})-(\ref{important_relation}) with $\gamma =1$. 
This representation provides us with a model of interacting spin-1/2 
fermions, which can be studied with the help of supersymmetry.

\subsection{Defining the Hilbert space}

To define the supercharges, we have first to fix the rules for determining  
the admissible configurations on a chosen  lattice. Here we will consider 
the case of a system of spin-1/2 fermions on a one-dimensional chain.

The construction of an $N=4$ supersymmetry structure for spin-1/2 
fermions is far from evident. We found that a straightforward 
generalization of the realization of $N=2$ supersymmetry 
for spin-less fermions, as presented in \cite{Fendley:2002sg}, fails
to reproduce the relation (\ref{important_relation}). To save
this relation, we have to define both the Hilbert space and the
supercharges in a quite evolved way.

The basic idea is  to incorporate the $SU(2)$ structure at the level of 
the definition of the Hilbert space. We consider then a system of  
spin-1/2 fermions on a chain, subject to the following conditions
\begin{itemize}
\item each site is occupied at most by one fermion,
\item among the  possible spin configurations associated to an
   even (odd) number of consecutive fermions, we allow only those which form 
   a singlet (doublet) of $SU(2)$.
\end{itemize}
This last condition is illustrated explicitly with the help of some
examples.

Consider first the case in which two fermions occupy neighboring sites. 
The only permitted configuration is the one in which the two fermions are 
in a spin singlet state ($\uparrow\downarrow -\downarrow\uparrow$). The 
other three possibilities ($\uparrow\uparrow,\downarrow\downarrow , 
\uparrow\downarrow+\downarrow\uparrow$) are excluded.

Note that the fully polarized version of the model we are building  
corresponds exactly to the $M_1$ model, as fermions of the same spin 
cannot occupy sites that are nearest neighbor on the chain. The supercharges
$\overline{Q}_{\uparrow}$ ($\overline{Q}_{\downarrow}$) and $Q_{\uparrow}$
($Q_{\downarrow}$) act  on the states with all the spins up (down)
in the same way as the supercharge in the $M_1$ model
\bea
&& \overline{Q}_{\uparrow} \left(\ldots \circ\circ\uparrow\circ\circ\ldots\right)
= \cdots+\left(\ldots\uparrow\circ\uparrow\circ\circ\ldots\right)-\left(\ldots\circ\circ\uparrow\circ\uparrow\ldots\right)+\cdots
\nonumber \\
&& Q_{\uparrow} \left(\ldots \circ\uparrow\circ\uparrow\circ\ldots \right)
= \cdots+\left(\ldots
  \circ\circ\circ\uparrow\circ\ldots\right)-\left(\ldots
  \circ\uparrow\circ\circ\circ\ldots\right)+\cdots \nonumber
 \eea
Here we represent with $\circ$ an empty state.

Consider now the case of three consecutive fermions. From the $SU(2)$ 
representation theory, the direct product of three spin $1/2$ representations
decomposes into a spin $3/2$ and two spin $1/2$ irreducible representations:
$1/2\otimes1/2\otimes1/2=3/2 \oplus 1/2 \oplus 1/2$. Among the $2^3=8$ 
possible configurations, we project out the states which participate in 
the spin $3/2$ representation, keeping only the states which
form a doublet ( $\uparrow\uparrow\downarrow+\downarrow\uparrow\uparrow-2
\uparrow\downarrow\uparrow$,
$\downarrow\downarrow\uparrow+\uparrow\downarrow\downarrow-2\,
 \downarrow\uparrow\downarrow$)  and
($\uparrow\uparrow\downarrow-\downarrow\uparrow\uparrow$,
$\downarrow\downarrow\uparrow-\uparrow\downarrow\downarrow$).

As we will show in the next section, the definition of the
supercharges depends on the global $SU(2)$ properties of a number of 
consecutive fermions and not on the particular single spin configuration. 
In the following we therefore always suppose to select one singlet (doublet) 
state among the possible ones formed by $2n$ ($2n+1$) consecutive fermions.
Here we introduce the following notation. We denote with $\ldots2\ldots$ the 
singlet formed by two consecutive fermions ($\ldots(\uparrow\downarrow
-\downarrow\uparrow)\ldots$), with $\ldots 3_{\alpha} \ldots$,
$\alpha=\uparrow,\downarrow$, the two states of a doublet formed by three 
consecutive fermions, etc. In the case of a cluster of three fermions, this  
notation will represent one of the two possible doublets, which can be chosen 
by following a certain criterion. For example,  $\ldots 3_{\uparrow}\ldots$ 
can indicate either the state 
$\ldots(\uparrow\uparrow\downarrow+\downarrow\uparrow\uparrow-2
 \uparrow\downarrow\uparrow)\ldots$ or the state  
$\ldots(\uparrow\uparrow\downarrow-\downarrow\uparrow\uparrow)\ldots$.
Starting from physical considerations, a possible criterion for choosing a 
particular singlet or doublet will be discussed later.

\begin{figure}
\centerline{
\epsfxsize=15cm
\epsfclipon
\epsfbox{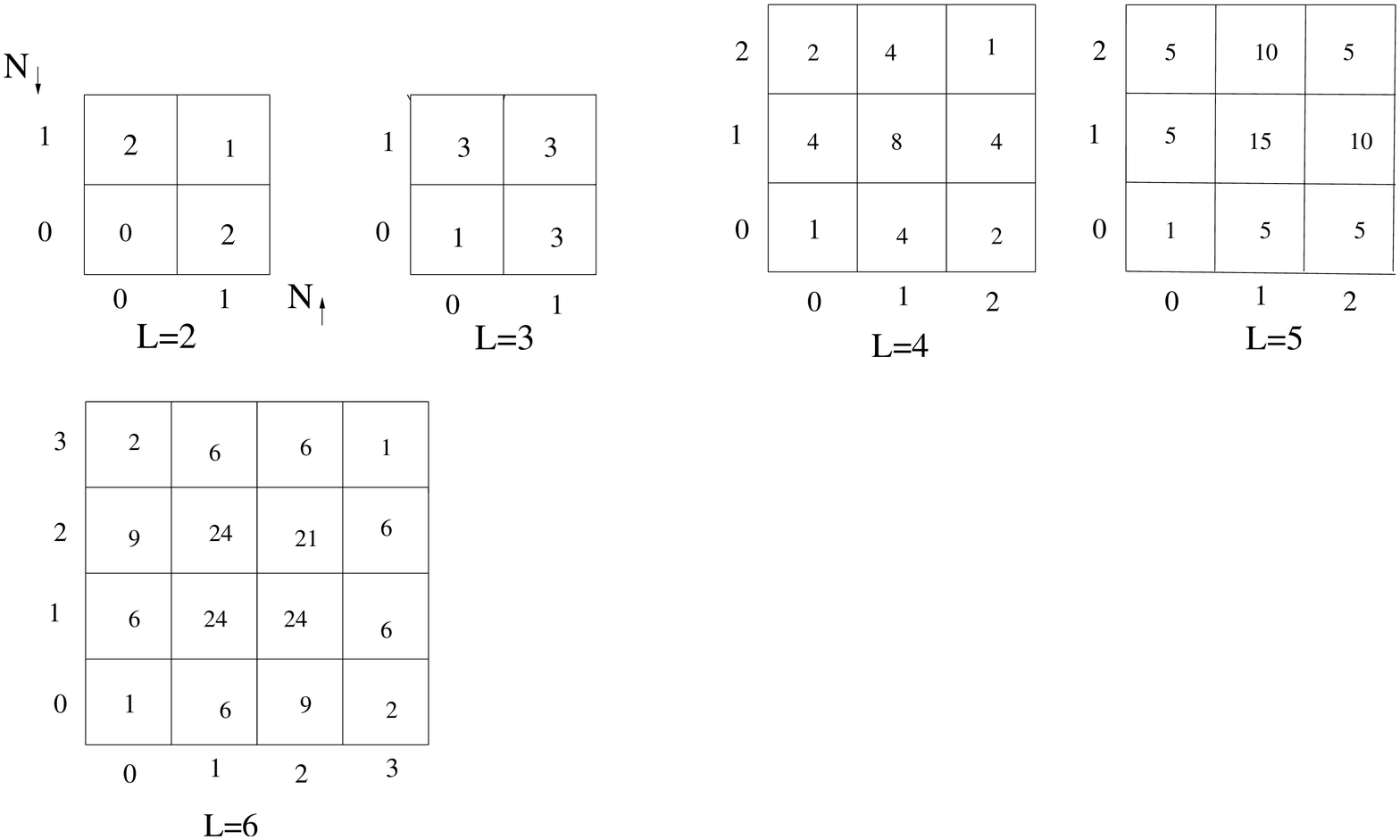}}
\caption{Table of the dimensions of the Hilbert space for each sector
  $(N_{\uparrow},N_{\downarrow})$ and for different chain lengths
  $L$. Here periodic boundary conditions are considered. }
\label{tavoladim}
\end{figure}

In general, denoting with $\ldots 2n\ldots$ and $\ldots (2n+1)_\alpha\ldots$  
a cluster of $2n$ and $2n+1$  consecutive fermions, we require that:
\beq
J^{a} \, 2n = 0 \ , \qquad 
J^{a} \, (2n+1)_{\alpha} = \sigma^a_{\alpha\beta}(2n+1)_{\beta} \ ,
\eeq
where $a=+,-,0$, $\alpha=\uparrow,\downarrow$ and
  $\sigma^a_{\alpha\beta}$ are the Pauli matrices.
A typical configuration reads
\beq
\ldots\circ\circ\circ 2 \circ 3_{\uparrow}\circ 3_{\downarrow}
\circ\uparrow\circ\circ\uparrow\circ 4\circ\circ\ldots
\nonumber
\eeq
This defines an Hilbert space $ \mathcal{H}$ with subspaces
${\mathcal{H}}_{N_{\uparrow},N_{\downarrow}}$ defined by the
condition that the eigenvalue of the spin $\alpha$ fermion number
operator $\hat{N}_{\alpha}$ be equal to $N_{\alpha}$. In 
Fig.~\ref{tavoladim} we give the dimension of the Hilbert space 
({\it i.e.}\ all the different configurations of clusters on the chain) 
for a chain of length $L=2,3,..,6$, when periodic boundary condition 
are taken.

\subsection{The supersymmetry generators}

\begin{figure}
\centerline{
\epsfxsize=10cm
\epsfclipon
\epsfbox{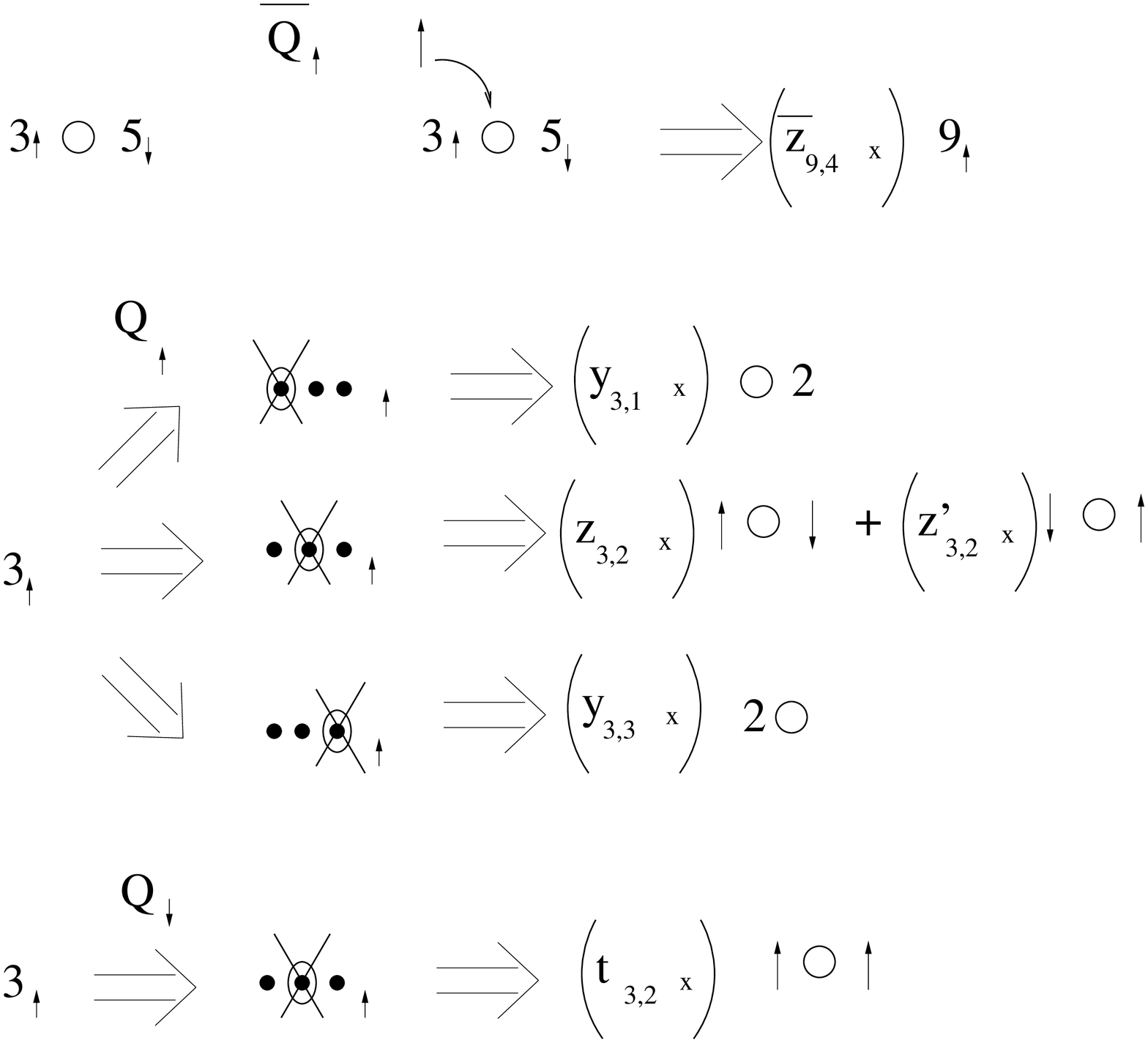}}
\caption{Examples of the action of the supersymmetry generators (with
  relative amplitudes) on the
restricted Hilbert space.}
\label{exdefQ}
\end{figure}

Instead of writing the supersymmetry operators in the second quantized
form, we prefer to list the non-vanishing amplitudes  they entail in
terms of fermions cluster, defined in the previous section. Before giving  
the complete construction of the supercharges operators, we
illustrate their action on some explicit examples shown in Fig.~\ref{exdefQ}

Consider two clusters of type $3_{\downarrow}$ and $5_{\uparrow}$ separated by
one empty site,$\ldots3_{\downarrow}\circ5_{\uparrow}\ldots$. The operator  
$\overline{Q}_{\uparrow}$ will fill the empty space between the two clusters with 
a fermion with spin $\uparrow$ and will create,  with a coefficient defined 
below, an unique  cluster of type $9_{\uparrow}$ , {\it i.e.}, 
$\overline{Q}_{\uparrow}\ldots3_{\downarrow}\circ 5_{\uparrow}\ldots\rightarrow
\ldots9_{\uparrow}\ldots $.

Suppose now that two consecutive clusters have the same spin, as in the 
configuration  $\ldots(2n+1)_{\uparrow}\circ(2m+1)_{\uparrow}\ldots$: 
the constraints on the Hilbert space imply that only a fermion with spin 
$\downarrow$  can be added, thus giving the possible amplitudes 
$ \overline{Q}_{\uparrow}\ldots (2n+1)_{\uparrow}\circ(2m+1)_{\uparrow}\ldots
\rightarrow 0$ and
$\overline{Q}_{\downarrow}\ldots(2n+1)_{\uparrow}\circ(2m+1)_{\uparrow}\ldots
\rightarrow \ldots(2n+2m+3)_{\uparrow}\ldots$.

The operators $Q_{\alpha}$ act in the opposite sense. Given a cluster of size
$n$, they take out a fermion with spin $\alpha$ at each position along the 
cluster, always respecting the conditions on the Hilbert space. Consider for 
example the  clusters $3_{\uparrow}$ and $3_{\downarrow}$. By applying 
the operator $Q_{\uparrow}$,  we find the states
$Q_{\uparrow}3_{\uparrow}\rightarrow\circ2,\downarrow\circ\uparrow,
\uparrow\circ\downarrow,2\circ$ and 
$Q_{\uparrow}3_{\downarrow}\rightarrow \downarrow\circ\downarrow$.

\begin{figure}
\centerline{
\epsfxsize=20cm
\epsfclipon
\epsfbox{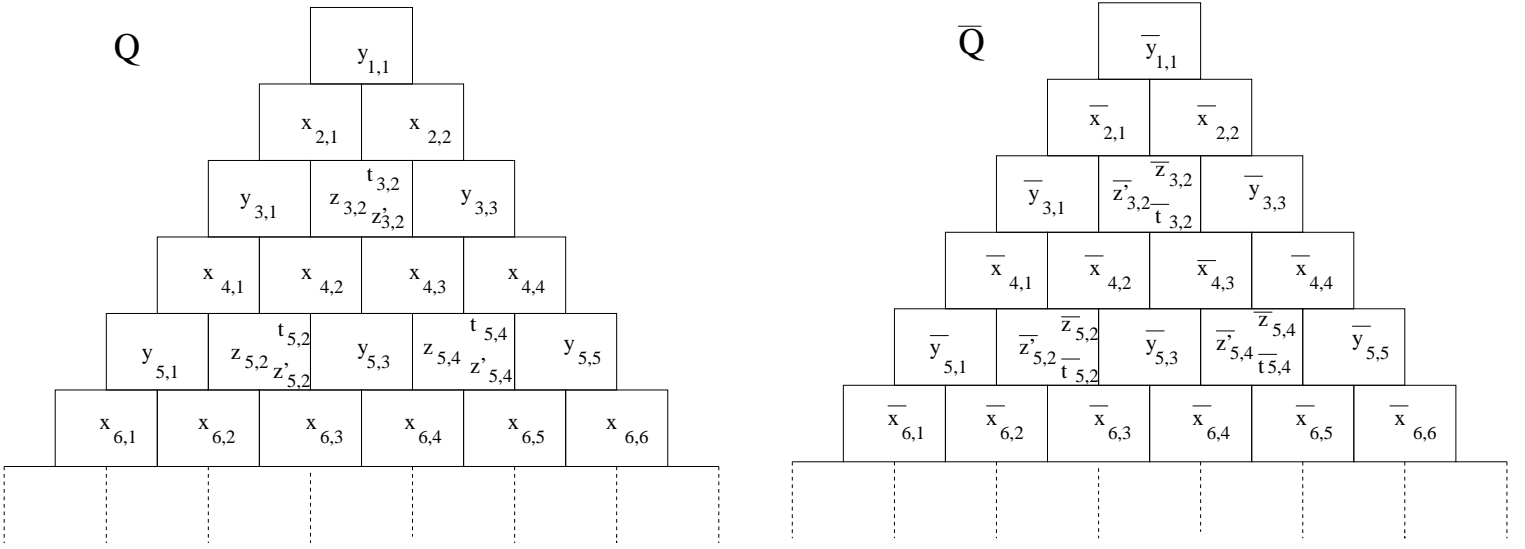}}
\caption{The amplitudes defining the supersymmetry generators can be listed in
  a pyramidal structure.}
\label{table_coeff}
\end{figure}

We  can now introduce in a compact form the amplitudes defining the 
supersymmetry operators $\overline{Q}_{\alpha}$
 \bea
\lefteqn{\overline{Q}_{\gamma} \ldots(2n+1)_{\alpha}\circ(2m+1)_{\beta}\ldots}
\nonumber \\
&=& (1-\delta_{\alpha,\beta})\left[\delta_{\gamma,\alpha}
     \bar{z}_{(2n+2m+3,2n+2),\gamma}+\delta_{\gamma,\beta}
     \bar{z}'_{(2n+2m+3,2n+2),\gamma}\right]
     \ldots(2n+2m+3)_{\gamma}\ldots
\nonumber \\
&& + \delta_{\alpha,\beta}(1-\delta_{\alpha,\gamma})\bar{t}_{(2n+2m+3,2n+2),
  \gamma}\ldots(2n+2m+3)_{\alpha}\ldots \ ,
\nonumber \\
\lefteqn{\overline{Q}_{\gamma} \ldots 2n\circ2m \ldots}
\nonumber \\ 
&=& \bar{y}_{(2n+2m+1,2n+1),\gamma}\ldots(2n+2m+1)_{\gamma}\ldots \ ,
\nonumber \\
\lefteqn{ \overline{Q}_{\gamma} \ldots2n\circ(2m+1)_{\alpha}\ldots} 
\nonumber\\
&=& (1-\delta_{\gamma,\alpha})\bar{x}_{(2n+2m+2,2n+1),\gamma}\ldots
    (2n+2m+2)\ldots \ ,
\nonumber \\
\lefteqn{\overline{Q}_{\gamma} \ldots(2n+1)_{\alpha}\circ2m\ldots}
\nonumber \\ 
&=& (1-\delta_{\gamma,\alpha})\bar{x}_{(2n+2m+2,2n+2),\gamma}
    \ldots(2n+2m+2)\ldots\label{Qdaga} \ ,
\eea 
and $Q_{\alpha}$ 
\bea
\lefteqn{Q_{\gamma} \ldots(2n+1)_{\alpha}\ldots}
\nonumber\\
&=& \delta_{\gamma,\alpha}\sum_{i=1,3,..}^{2n+1}
    y_{(2n+1,i),\gamma}\ldots(2n+1-i)\circ(i-1)\ldots
\nonumber \\
&& +\delta_{\gamma,\alpha}(1-\delta_{\gamma,\beta})\sum_{i=2,4,..}^{2n}
  z_{(2n+1,i),\gamma}\ldots(2n+1-i)_{\alpha}\circ(i-1)_{\beta}\ldots
\nonumber \\
&& +\delta_{\gamma,\alpha}(1-\delta_{\gamma,\beta})\sum_{i=2,4,..}^{2n}
  z'_{(2n+1,i),\gamma}\ldots(2n+1-i)_{\beta}\circ(i-1)_{\alpha}\ldots
\nonumber \\
&& +(1-\delta_{\gamma,\alpha})\sum_{i=2,4,..}^{2n}t_{(2n+1,i),\gamma}
   \ldots(2n+1-i)_{\alpha}\circ (i-1)_{\alpha}\ldots \ ,
\nonumber \\
\lefteqn{Q_{\gamma} \ldots2n\ldots} 
\nonumber\\
&=& (1-\delta_{\gamma,\alpha})\left[\sum_{i=1,3,..}^{2n-1}
     x_{(2n,i),\gamma}\ldots(2n-i)_{\alpha}\circ(i-1)\ldots
     +\sum_{i=2,4,..}^{2n} x_{(2n,i),\gamma}\ldots(2n-i)\circ(i-1)_{\alpha}
     \ldots\right] .
\nonumber\\
\label{Q}
\eea
In Fig.~\ref{exdefQ} we show some of these matrix elements. The whole
set of coefficients defining the supersymmetry generators  
can be displayed  as in Fig.~\ref{table_coeff}.

The values of these coefficients are  determined by imposing the algebra 
(\ref{comm1})-(\ref{important_relation}). The procedure goes as follows.
We consider, for example, the operators
$\overline{Q}_{\uparrow}$ and $Q_{\downarrow}$ and we determine the 
coefficients in Fig.~\ref{table_coeff} row by row starting from the
top of the pyramid. The conditions of nil-potency of the supersymmetry 
operators, $\overline{Q}^2_{\uparrow}=0$ and $Q^2_{\downarrow}=0$,
lead to a series of recursions of the type:
\bea
X_{(n,m),\alpha}X_{(q,n+1),\alpha}
&=& X_{(q-m,n+1-m),\alpha}X_{(q,m),\alpha}
\nonumber \\
\bar{X}_{(n,m),\alpha}\bar{X}_{(q,n+1),\alpha}
&=& \bar{X}_{(q-m,n+1-m),\alpha}\bar{X}_{(q,m),\alpha},
\eea
where $X_{(a,b),\alpha}$ denotes one of the possible type of amplitude
and $m\leq n+1\leq q$.  The above recursions  leave some free
parameters which are then tuned to satisfy  the further relation 
$\{\overline{Q}_{\uparrow},Q_{\downarrow}\}=-J^{+}$. Modulo an overall
 normalization of the supercharges, $\overline{Q}_{\uparrow}\to
N\times\overline{Q}_{\uparrow}$ and 
$Q_{\downarrow}\to N^{-1}\times Q_{\downarrow}$ (we will show later
 that the norm of the cluster
states $2n$ and $(2n+1)_{\alpha}$ depends on the choice of this
normalization), all the possible parameters
 are then determined.

\begin{figure}
\centerline{
\epsfxsize=20cm
\epsfclipon
\epsfbox{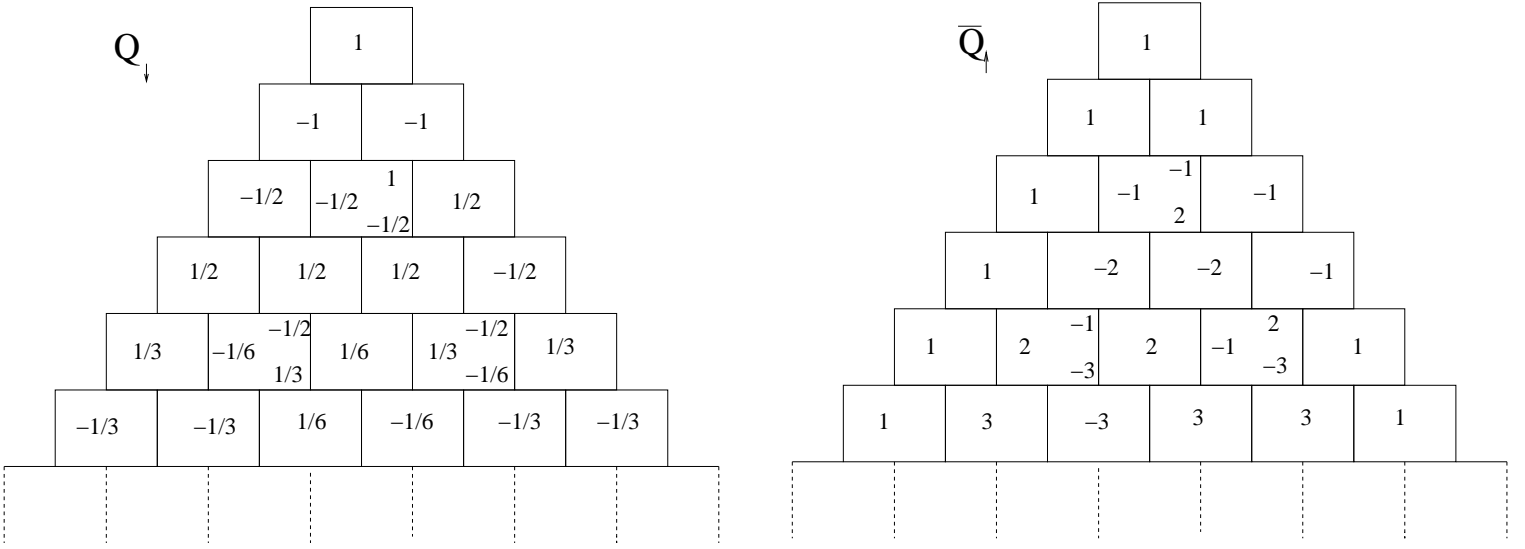}}
\caption{Values of the amplitudes of the supersymmetry generators fixed by
  imposing the algebra (\ref{comm1})-(\ref{important_relation}).}
\label{tavola_num}
\end{figure}

The values found in this way are shown in Fig.~\ref{tavola_num}, where a
particular normalization has been chosen. 
It is important to remark that, already at this stage, the number
of constraints imposed by the algebra is larger than the number
of adjustable parameters $\bar{X}_{a,b}$. This is the reason why, for
example, we cannot adjust the parameters to realize the algebra
(\ref{important_relation}) with $\gamma=-1$ or $\gamma=0$.

The next step is to find the coefficients of the other two operators,
$\overline{Q}_{\downarrow}$ and $Q_{\uparrow}$.  The conditions
 $\{\overline{Q}_{\uparrow},\overline{Q}_{\downarrow}\}=0$ and
$\{Q_{\uparrow},Q_{\downarrow}\}=0$ (see (\ref{comm1})) yield
the relations between the up and down sectors
\bea
&& \bar{X}_{(2n+1,2m),\uparrow}=\bar{X}_{(2n+1,2m),\downarrow} \ ,
\qquad
X_{(2n+1,2m),\uparrow}=X_{(2n+1,2m),\downarrow} 
\qquad \mbox{for} \ X=t,z,z' \ , 
\nonumber\\
&& \bar{y}_{(2n+1,2m+1),\uparrow} = (-1)^{n}\bar{y}_{(2n+1,2m+1),\downarrow} 
\ ,
\qquad
y_{(2n+1,2m+1),\uparrow}=(-1)^{n}y_{(2n+1,2m+1),\downarrow} \ ,
\nonumber\\
&& \bar{x}_{(2n,2m+1),\uparrow} = (-1)^{n+m}\bar{x}_{(2n,2m+1),\downarrow} \ ,
\qquad 
\bar{x}_{(2n,2m),\uparrow}=(-1)^{m}\bar{x}_{(2n,2m),\downarrow} \ ,
\nonumber \\
&& x_{(2n,2m+1),\uparrow}=(-1)^{n+m}x_{(2n,2m+1),\downarrow} \ ,
\qquad
x_{(2n,2m),\uparrow}=(-1)^{m}x_{(2n,2m),\downarrow} \ .
\label{sign_rule}
\eea
The above sign rules can be nicely explained in terms
of the parity  of the singlets under a spin reversal. In particular, 
if a string of $2n$ fermions takes a sign $(-1)^n$ under a spin reversal 
operation, then the equations (\ref{sign_rule}) are easily obtained. 
Note that a singlet formed by two fermions,
$\uparrow\downarrow-\downarrow\uparrow$ has an odd parity, and the
two singlets of four consecutive fermions,
$\uparrow\uparrow\downarrow\downarrow+\uparrow\downarrow\downarrow\uparrow
+\downarrow\downarrow\uparrow\uparrow+\downarrow\uparrow\uparrow\downarrow
-2\uparrow\downarrow\uparrow\downarrow-2\downarrow\uparrow\downarrow\uparrow$
and  
$\uparrow\uparrow\downarrow\downarrow-\uparrow\downarrow\downarrow\uparrow+
\downarrow\downarrow\uparrow\uparrow-\downarrow\uparrow\uparrow\downarrow$, 
have an even parity. 

Choosing a normalization such that
$\bar{y}_{(2n+1,1),\uparrow}=\bar{x}_{(2n,1),\uparrow}=1$, 
the general solution for the coefficients reads as follows:
\bea
\overline{Q}_{\uparrow}&&\left\{
\begin{array}{ll}
  \bar{t}_{(2n+1,2m),\uparrow}=(-1)^{mn+1}\left(\begin{array}{ccc}n+1
  \\ m\end{array}\right) \ , &
  \bar{y}_{(2n+1,2m+1),\uparrow}=(-1)^{nm}\left(\begin{array}{ccc}n
  \\ m\end{array}\right) \ ,
  \nonumber \\
  \bar{z}_{(2n+1,2m),\uparrow}=(-1)^{(m-1)(n-1)+1}\left(\begin{array}{ccc}n
  \\ m-1\end{array}\right) \ , & 
  \bar{z}'_{(2n+1,2m),\uparrow}=(-1)^{m(n-1)+1}
  \left(\begin{array}{ccc}n \\ m\end{array}\right) \ ,
  \nonumber\\
  \bar{x}_{(2n,2m),\uparrow}=(-1)^{m(n-1)}\left(\begin{array}{ccc}n
  \\ m\end{array}\right) \ , &
  \bar{x}_{(2n,2m+1),\uparrow}=(-1)^{mn}\left(\begin{array}{ccc}n
  \\ m\end{array}\right) \ ,
\end{array} \right.
\\
Q_{\uparrow}&&\left\{
\begin{array}{ll}
  t_{(2n+1,2m),\uparrow}=\frac{(-1)^{mn+1}}{m}\left(\begin{array}{ccc}n
  \\ m\end{array}\right)^{-1} \ , & 
  y_{(2n+1,2m+1),\uparrow}=\frac{(-1)^{nm}}{m+1}\left(\begin{array}{ccc}
  n+1 \\ m+1\end{array}\right)^{-1} \ ,
  \nonumber \\
  z_{(2n+1,2m),\uparrow}=\frac{(-1)^{(m-1)(n-1)+1}}{m+1}\left(\begin{array}{ccc}n+1
  \\ m+1\end{array}\right)^{-1} \ , & 
  z'_{(2n+1,2m),\uparrow}=\frac{(-1)^{m(n-1)+1}}{m}\left(\begin{array}{ccc}n+1
  \\ m\end{array}\right)^{-1} \ ,
  \nonumber\\
  x_{(2n,2m),\uparrow}=\frac{(-1)^{m(n-1)}}{m}\left(\begin{array}{ccc}n
  \\ m\end{array}\right)^{-1} \ , &  
  x_{(2n,2m+1),\uparrow}=\frac{(-1)^{mn}}{m+1}\left(\begin{array}{ccc}n
  \\ m+1\end{array}\right)^{-1} \ .
\end{array} \right.
\eea
From the above formulas, the amplitudes associated to
$\overline{Q}_{\downarrow}$ and $Q_{\downarrow}$ are easily obtained by using 
the relations (\ref{sign_rule}).

The norms $N_{2n}\doteq<2n|2n>$ and 
$N_{2n+1}\doteq<(2n+1)_{\alpha}|(2n+1)_{\alpha}>$ of the cluster states 
$2n$ and $(2n+1)$ are fixed by using the fact the operators 
$\overline{Q}_{\alpha}$ and $Q_{\alpha}$ are self conjugate. For example, the
condition
\beq
<2n|\overline{Q}_{\uparrow}|(2n-2q-1)_{\downarrow}\circ2q>
=<(2n-2q-1)_{\downarrow}\circ2q|Q_{\uparrow}|2n>
\eeq
yields the following relation between the norms and the amplitudes
associated to the supersymmetry generators
\beq
\bar{x}_{(2n,2n-2q),\uparrow}N_{2n}
=x_{(2n,2n-2q),\uparrow}N_{2n-2q-1}N_{2q} \ .
\label{conj_cond}
\eeq
Given as initial conditions $N_0=N_1=1$, one finds that the  
above condition is satisfied if
\beq
N_{2n}=\frac{1}{(n!)^2}\quad \mbox{and} \quad N_{2n+1}=\frac{1}{(n+1)!n!} \ .
\label{norm}
\eeq
One can verify that all the other self-conjugacy relations like the eq. 
(\ref{conj_cond}) are all respected by taking the norms (\ref{norm}),
thus confirming the consistency of all the construction.
Normalizing the basis states as  
$|2n>\to N_{2n}^{-1/2}|2n>$ and  $|(2n+1)_{\alpha}>\to N_{2n+1}^{-1/2}|(2n+1)_{\alpha}>$, 
the amplitudes of the supercharges operators take a very simple form:
\bea
&& \bar{t}_{(2n+1,2m),\uparrow} = t_{(2n+1,2m),\uparrow}=(-1)^{mn+1}
\sqrt{\frac{(n+1)(n-m)}{m}} \ ,
\nonumber \\
&& \bar{y}_{(2n+1,2m+1),\uparrow} = y_{(2n+1,2m+1),\uparrow}
=(-1)^{nm}\sqrt{\frac{1}{n+1}} \ ,
\nonumber \\
&& \bar{z}'_{(2n+1,2m),\uparrow}= z'_{(2n+1,2m),\uparrow}
=(-1)^{m(n-1)+1}\sqrt{\frac{m}{(n+1)(n-m+1)}} \ ,
\nonumber \\
&& \bar{z}_{(2n+1,2m),\uparrow} 
   = z_{(2n+1,2m),\uparrow}=(-1)^{(m-1)(n-1)+1}\sqrt{\frac{n-m+1}{m(n+1)}} \ ,
\nonumber\\
&& \bar{x}_{(2n,2m),\uparrow} =  x_{(2n,2m),\uparrow}
   = (-1)^{m(n-1)}\sqrt{\frac{1}{m}} \ ,
\nonumber \\
&& \bar{x}_{(2n,2m+1),\uparrow} = x_{(2n,2m+1),\uparrow}
   =(-1)^{mn}\sqrt{\frac{1}{n-m}} \ .
\eea
With this last normalization, the operators $\overline{Q}_{\alpha}$ and
$Q_{\alpha}$ conjugate w.r.t.\ one another. In the examples presented
in section~2.4, we shall denote normalized basis states with
brackets, $\langle \ldots \rangle$.

Having constructed completely the supersymmetry operators, one can
verify that they realize the algebra (\ref{comm1})-(\ref{important_relation}). 
We have thus obtained a non-linear realization of the $N=4$ supersymmetry
algebra (\ref{comm1})-(\ref{important_relation}) at $\gamma=1$, on
a Hilbert space built from spin-1/2 fermions. The corresponding Hamiltonian
and supersymmetric ground states are discussed in the next section.

\subsection{Hamiltonian and supersymmetric ground states}
\label{model}

The Hamiltonian (\ref{hamiltonian}) provided by the supercharges
(\ref{Qdaga}) and (\ref{Q})  acts on the Hilbert space
$\mathcal{H}$ composed by  clusters of electrons forming singlet
or doublet of total spin, all spaced by one or more unoccupied
sites. The amplitudes associated to the terms of the Hamiltonian  
listed below are given in the Appendix.
The action of the Hamiltonian can be divided into
\begin{itemize}
 \item split-join terms. Two clusters spaced by one
 site split and join to form all the possible pairs of
 clusters that
  preserve the number of fermions with spin $\uparrow$ and
  $\downarrow$ and respect  the constraints on the Hilbert space: 
\begin{itemize}
    \item     $2n\circ2m\leftrightarrow (1-\delta_{\alpha,\beta})
(2n+2m-2q-1)_{\alpha}\circ(2q+1)_{\beta}$  with
$\alpha,\beta=\uparrow,\downarrow$ and  $q=0,1,..,n+m-1$.
\item $2n\circ2m\leftrightarrow (2n+2m-2q)\circ2q$  with
$q=0,1..,m-1$,
\item $2n\circ(2m+1)_{\alpha}\leftrightarrow
(2n+2m-2q+1)_{\alpha}\circ2q$ with $q=0,1..,n+m$,
\item
  $2n\circ(2m+1)_{\alpha}\leftrightarrow(2n+2m-2q)\circ(2q+1)_{\alpha}$
  with $q=0,..,n+m-1$.
\end{itemize}
Among these amplitudes we find the hopping terms where a
single cluster is moved by one position: $2n\circ\leftrightarrow
2n\circ$ and $(2n+1)_{\alpha}\circ\leftrightarrow
(2n+1)_{\alpha}\circ$ .
\item potential terms (diagonal terms):
\begin{itemize}  
\item $2n\leftrightarrow
2n$,
\item  $(2n+1)_{\alpha}\leftrightarrow
(2n+1)_{\alpha}$.
\end{itemize}

\end{itemize}
Some explicit examples are given
 below:
\bea
H \langle \circ\uparrow\circ\downarrow\circ \rangle
&=&
 5 \langle \circ\uparrow\circ\downarrow\circ \rangle
 +2\left[ \langle \uparrow\circ\circ\downarrow\circ \rangle
 + \langle \circ\uparrow\circ\circ\downarrow \rangle \right]
 + \langle \circ2\circ\circ \rangle
 + \langle \circ\circ2\circ \rangle
 + \langle \circ\downarrow\circ\uparrow\circ \rangle \ ,
 \nonumber \\
H \langle \circ4\circ \rangle &=&
 \frac{22}{3} \langle \circ4\circ \rangle
 +\frac{2 \sqrt{2}}{3}\left[ \langle \circ3_{\uparrow}\circ\downarrow \rangle
                   + \langle \circ3_{\downarrow}\circ\uparrow \rangle\right]
 -\frac{\sqrt{2}}{3}\left[ \langle 3_{\uparrow}\circ\downarrow\circ \rangle
                   + \langle 3_{\downarrow}\circ\uparrow\circ \rangle\right]
 \nonumber\\
&&
 -\frac{\sqrt{2}}{3}\left[ \langle \circ\uparrow\circ3_{\downarrow}\rangle
                   + \langle \circ\downarrow\circ3_{\uparrow}\rangle\right]
 +\frac{2\sqrt{2}}{3}\left[ \langle \uparrow\circ3_{\downarrow}\circ\rangle
                   + \langle \downarrow\circ3_{\uparrow}\circ\rangle\right]
 -\frac{1}{3}\left[\langle 4\circ\circ\rangle+ \langle \circ\circ4\rangle
                   \right]
 \nonumber\\
&&
 -\frac{4}{3}\left[\langle \circ2\circ2 \rangle
                   + \langle 2\circ2\circ\rangle\right] \ .
\eea

\begin{figure}
\centerline{
\epsfxsize=7cm
\epsfclipon
\epsfbox{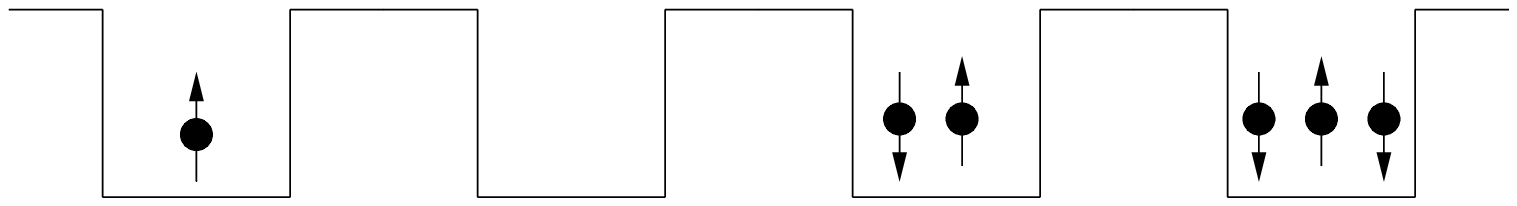}}
\caption{Spin $1/2$ particles confined in 
a one dimensional periodic potential.}
\label{mapp1}
\end{figure}

\begin{figure}
\centerline{
\epsfxsize=17cm
\epsfclipon
\epsfbox{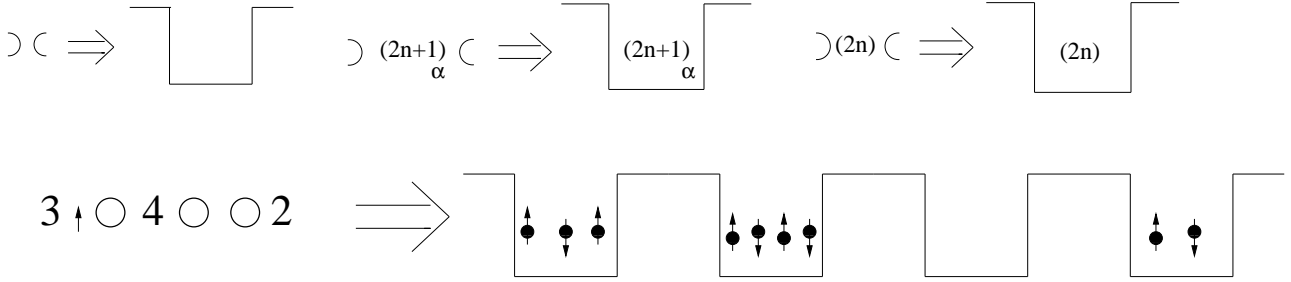}}
\caption{The model of spin $1/2$ fermions on a chain is mapped into a 
         model of spin $1/2$ fermions confined to a one-dimensional periodic 
         potential.}
\label{mapp2}
\end{figure}

\begin{figure}
\centerline{
\epsfxsize=10cm
\epsfclipon
\epsfbox{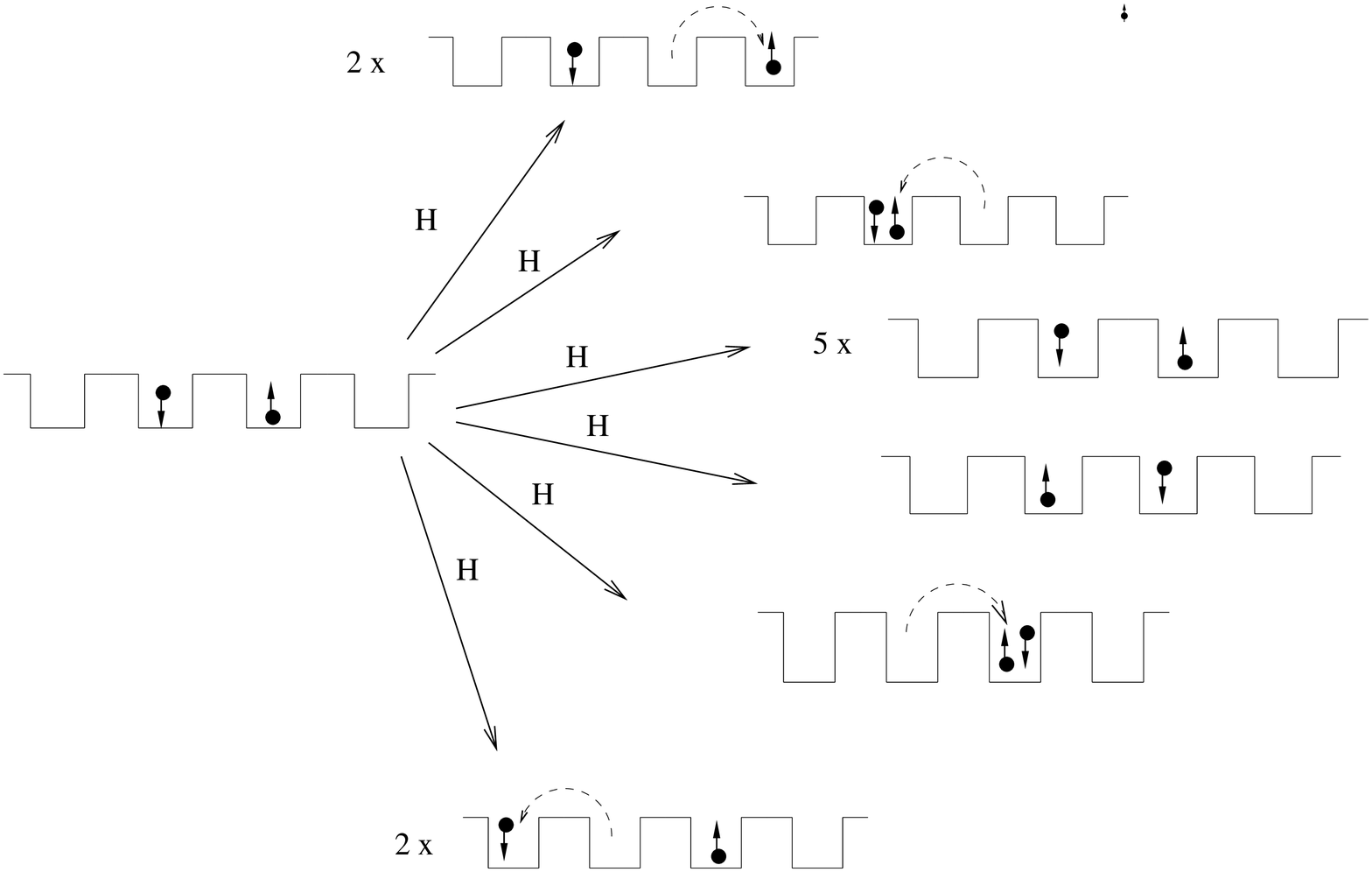}}
\caption{Transfer of particles between adjacent wells as induced by
  the supersymmetric Hamiltonian. }
\label{mapp3}
\end{figure}

A possible interpretation of our supersymmetric model is the following.
Consider a system of spin $1/2$ particles subjected to a one dimensional 
periodic potential as shown in Fig.~\ref{mapp1}. 
In the mapping given in Fig.~\ref{mapp2}, a cluster of $2n$ ($2n+1$) fermions 
represent a number $2n$ ($2n+1$) of spin $1/2$ particles confined in a well 
and forming a singlet (doublet) state of total spin. 
This state could originate from an anti-ferromagnetic interaction between 
these particles. We can for example suppose that the particles confined in
a well form the ground state of an anti-ferromagnetic Heisenberg Hamiltonian. 
Actually, this hypothesis would provide us with a reasonable criterion 
to select one singlet (doublet) among the possible ones formed by $2n$
($2n+1$) spin $1/2$ particles. We could imagine this system has
two characteristic times, $\tau_1$, $\tau_2$, with $\tau_1\ll\tau_2$. In a 
time of order of $\tau_1$, the particles arrange themselves to form the 
ground state of an anti-ferromagnetic spin energy operator. From this point 
of view, the Hamiltonian we have defined describes a particular dynamics 
induced by the transfer of particles between adjacent wells (see 
Fig.~\ref{mapp3}), which takes place over times of the order of $\tau_2$.

We consider now the case of a chain of finite length $L$. It is important 
to remember here that to realize the algebra 
(\ref{comm1})-(\ref{important_relation}) with $\gamma=1$ we need to allow 
for strings of consecutive fermions of any size. The construction of the 
supersymmetry generators as given in the previous section is thus consistent 
only in the case of an infinite chain. If the chain is finite, specific 
boundary conditions tend to be in conflict with the supersymmetry 
algebra in the subspaces ${\mathcal{H}}_{N_{\uparrow},N_{\downarrow}}$ 
with $N_{\uparrow}+N_{\downarrow}\geq L-1$, as the notions of 
empty sides on the left and right of a length $(L-1)$ string
become ambiguous.

\begin{figure}
\centerline{
\epsfxsize=15cm
\epsfclipon
\epsfbox{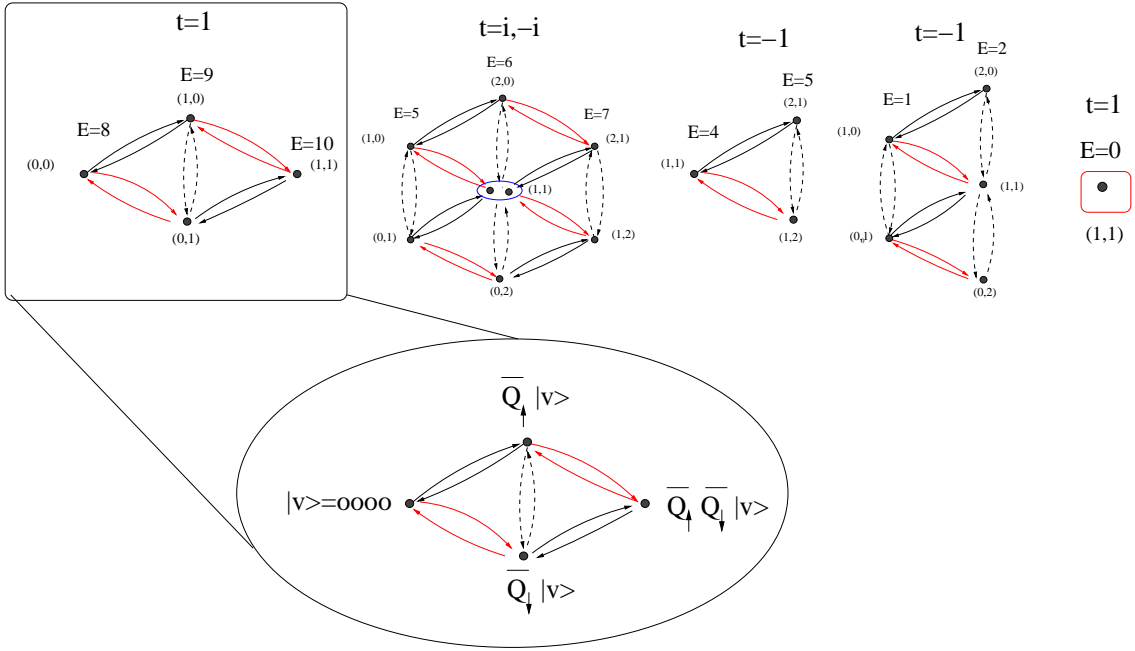}}
\caption{Spectrum of the model on a chain of length $L=4$. The
  supermultiplet formed by acting with supersymmetry operators on the vacuum
  state is shown in detail.}
\label{L4table}
\end{figure}

The first example we consider is the case $L=4$, with periodic boundary 
conditions. The corresponding dimensions of the Hilbert spaces 
${\mathcal{H}}_{N_{\uparrow},N_{\downarrow}}$ are given in 
Fig.~\ref{tavoladim}. We have computed the eigenvalues of 
the Hamiltonian $H$ in each sector ${\mathcal{H}}_{N_{\uparrow},
N_{\downarrow}}$ with $N_{\uparrow}+N_{\downarrow}<3$, thus
avoiding the sectors $(1,2)$, $(2,1)$ and $(2,2)$ where, due
to the periodic boundary conditions, the action of the
supercharges becomes ambiguous. We have also kept track of the 
multiplet structure under the supersymmetry algebra, and of
the eigenvalues $t$ of the action of the translation operator
(satisfying, in general, $t^L=1$). The complete results are
given in Fig.~\ref{L4table}. To complete the multiplets at $t=-1,i,-i$ 
that involve states at fermion numbers $(2,1)$ and $(1,2)$, we have 
assumed that the supercharges $\overline{Q}_{\uparrow}$ and 
$\overline{Q}_{\downarrow}$ annihilate these states.

It is important to stress the existence of a unique ground
state, $|gs>_{4}$  with energy $E=0$, and  thus
$\overline{Q}_{\alpha}|gs>_4=Q_{\alpha}|gs>_4=0$, in the sector $(1,1)$.
Its explicit form in terms of clusters is:
\beq
|gs>_{4}=\left[ \langle \uparrow\circ\downarrow\circ\rangle
                 - \langle \downarrow\circ\uparrow\circ\rangle
            + \langle \circ\uparrow\circ\downarrow\rangle
            - \langle \circ\downarrow\circ\uparrow\rangle\right]
        -\left[\langle 2\circ\circ\rangle
               + \langle\circ2\circ\rangle
               + \langle\circ\circ2\rangle
               + \langle2\circ\circ2\rangle\right],
\eeq
where the notation $\langle 2\circ\circ 2\rangle$ indicates that the 
fermions at the border
of the chain  form a singlet (this is due to the period boundary conditions). 
 
\begin{figure}
\centerline{
\epsfxsize=15cm
\epsfclipon
\epsfbox{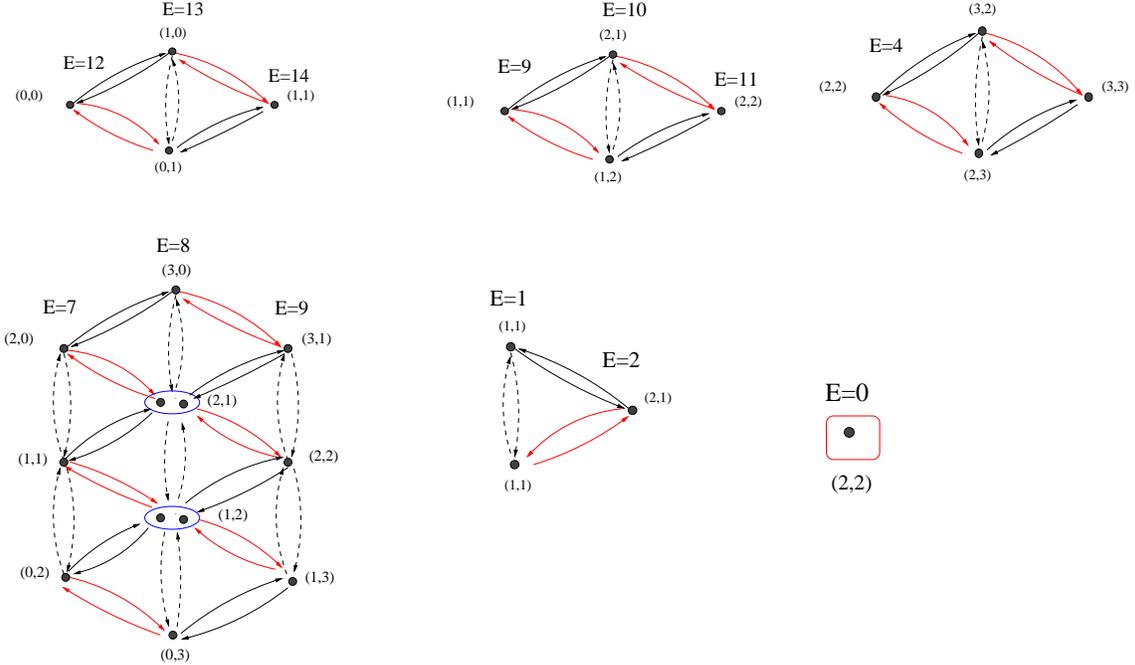}}
\caption{Supermultiplets formed by the translationally invariant
  eigenstates for a chain of length $L=6$.}
\label{L6table}
\end{figure}

We have obtained further analytic results  for chains of length
$L=5,6,7,8$ with periodic boundary conditions. In particular we were
interested in the eventual presence of supersymmetric ground states  
like the one appearing in the case $L=4$. When  $L=5$ or $L=7$, there is
no state $|gs>$ which satisfies $\overline{Q}_{\alpha}|gs>=Q_{\alpha}|gs>=0$. 
On the contrary, we have found such states in the cases $L=6$ and $L=8$.
For $L=6$, the supermultiplets formed by the $t=1$ eigenvectors are
shown in Fig.~\ref{L6table}. We notice the presence of a  
unique supersymmetric ground state $|gs>_6$ in the sector $(1,1)$, 
with the following form
\beq
|gs>_6 = 4\left[\langle\uparrow\circ\circ\downarrow\circ\circ\rangle-
              \langle\downarrow\circ\circ\uparrow\circ\circ\rangle\right]
        +3\left[\langle\uparrow\circ\downarrow\circ\circ\circ\rangle-
              \langle\downarrow\circ\uparrow\circ\circ\circ\rangle\right]
        +\left[\langle2\circ\circ\circ\circ\rangle\right]+.. \ ,
\eeq
with the $\ldots$ representing all possible translations of the
states shown.
In the case $L=8$, we have found a unique ground state $|gs>_8$  in the 
sector $(2,2)$ with translational eigenvalue $t=1$. The corresponding 
expression is
\bea
|gs>_8 &=& 
  4\sqrt{2}\left[\langle 3_{\uparrow}\circ\circ\downarrow\circ\circ\rangle
         -\langle 3_{\downarrow}\circ\circ\uparrow\circ\circ\rangle\right]
  \nonumber\\
&&
  - 3\sqrt{2} \left[ \langle 3_{\uparrow}\circ\downarrow\circ\circ\circ\rangle
          -\langle 3_{\downarrow}\circ\uparrow\circ\circ\circ\rangle\right]
  - 3\sqrt{2} \left[ \langle\uparrow\circ3_{\downarrow}\circ\circ\circ\rangle
           -\langle\downarrow\circ3_{\uparrow}\circ\circ\circ\rangle\right]
   \nonumber\\
&&
   + 4 \langle 2\circ\circ 2\circ\circ \rangle
   + \langle 4\circ\circ\circ\circ \rangle  + \ldots \nonumber.
\eea
Again, the $\ldots$ complete the expression to one that is 
translationally invariant. 

Numerical diagonalization of the Hamiltonian up to $L=14$ sites
gave the following results.While for $L=9$, $L=11$ and $L=13$ 
there are no supersymmetric ground states, such states do exist for
$L=10$, $L=12$ and $L=14$, in the sectors with fermion 
numbers $(3,3)$, $(4,4)$ and $(5,5)$, respectively.  

At present, we do not have a good grasp of the pattern for the
existence of supersymmetric ground states for general $L$. The
quick arguments based on the Witten index, which guarantee
the existence of such states in many $N=2$ supersymmetric models, 
do not apply in these $N=4$ supersymmetric models. From our explicit
results up to $L=14$, there is an obvious conjecture that supersymmetric 
ground states exist for general even $L$.

While, again, we lack a solid argument for determining the 
fermion number of these putative ground states, we observe the
following. Focusing on a length-$n$ string of consecutive 
fermions, one quickly finds that the potential energy $W_c(n)$ has a 
logarithmic dependence on $n$, $W_c(n)\sim \ln{n}$. In contrast, 
the contribution $W_h(n)$ to the potential energy from a string of $n$ 
consecutive empty sites is linear in $n$, $W_h(n)\sim 2n$. Thus the
system clearly favors the formation of large clusters. If indeed 
supersymmetric ground states exist for general even $L$, we expect them
at a number of holes that either remains at a finite value or  
grows logarithmically with $L$ (in this case 
the supersymmetric ground state will be in the sector $(n,n)$ with 
$L-2n \propto \ln{L}$).

Another important issue which we leave unsettled is that of conformal
invariance. It will be interesting to determine if the model is critical,
{\it i.e.}, if the excitation energies above the ground state decrease as 
$1/L$ in the continuum limit. If this is the case, one expects that
contact can be made with a form of conformal field theory.

\section{Conclusion}
In this concluding section we reiterate some of the remarkable 
properties of the model we considered, and we make some comments
on related issues.

We have introduced a model of interacting spin-$1/2$ fermions on a chain 
with a manifest $SU(2)$ extended $N=4$ supersymmetry. Our representation
of $N=4$ supersymmetry is highly non-linear, as it is entirely built
from degrees of freedom that are fermionic. We have looked for a
supersymmetric model where $SU(2)$ spin symmetry is faithfully 
represented, and this has led us to a somewhat unusual restricted Hilbert 
space, with anti-ferromagnetic correlations built in from the start.
The algebraic structure we have uncovered is very rich, but we
are lacking a systematic mathematical framework. Such a framework
will be most valuable, as it will allow us to further work out our 
present model and to decide on possibilities for alternative realizations 
of $N=4$ supersymmetry.

We have found supersymmetric (zero energy) ground states for even $L$
up to $L=14$. In physical terms, they represent a (small) number of
holes in an anti-ferromagnetic background. Our concrete realization
of supersymmetry is restricted to one spatial dimension, but the
general idea of exploiting supersymmetry is not. [See 
ref.~\cite{Fendley:2005} where ground state properties for spin-less
fermions on a variety of $D=2$ lattices are presented.] There is thus 
a possibility for exploiting supersymmetry in the analysis of doped 
antiferromagnets on $D=2,3$ lattices. If this can be made to work, it
is a potentially potent tool, which may supplement recent developments,
where important progress on RVB states in $D=2$ antiferromagnets was
made with the help of the analysis of associated quantum dimer models
\cite{Raman:2005}.

\vskip 1cm

\noindent \underline{Acknowledgments}

We wish to thank P.~Fendley for valuable discussions. This
research is supported by a grant from the Netherlands
Organization of Scientific Research (NWO). 
  
\newpage

\begin{appendix}
\section{Amplitudes associated to the Hamiltonian}
We give in this appendix the matrix elements defining the Hamiltonian
(see section \ref{model}) in terms of the amplitudes of the supercharges.

\noindent \underline{split-join terms}
\begin{itemize}
\item     
$(2n)\circ(2m)\leftrightarrow (1-\delta_{\alpha,\beta})
(2n+2m-2q-1)_{\alpha}\circ(2q+1)_{\beta}$, with $q=0,1,..,n+m-1$:
\bea
\alpha,\beta=\uparrow,\downarrow
&\!\!\to\!\!&\sum_{\gamma}z_{(2n+2m+1,2n+2m-2q),\gamma}\bar{y}_{(2n+2m+1,2n+1),\gamma}+\bar{y}_{(2n+2m-2q-1,2n+1),\uparrow}x_{(2m,2m-2q-1),\uparrow}\nonumber\\
\alpha,\beta=\downarrow,\uparrow
&\!\!\to\!\!&\sum_{\gamma}z'_{(2n+2m+1,2n+2m-2q),\gamma}\bar{y}_{(2n+2m+1,2n+1),\gamma}+\bar{y}_{(2n+2m-2q-1,2n+1),\downarrow}x_{(2m,2m-2q-1),\downarrow}
\nonumber
\eea
for $q=0,1..,m-1$ and 
\bea
\alpha, \beta=\uparrow,\downarrow
&\!\!\to\!\!&\sum_{\gamma}z_{(2n+2m+1,2n+2m-2q),\gamma}\bar{y}_{(2n+2m+1,2n+1),\gamma}+\bar{y}_{(2m+2k+1,2k+1),\downarrow}x_{(2n,2n-2k),\downarrow}\nonumber\\
\alpha, \beta=\downarrow,\uparrow
&\!\!\to\!\!&\sum_{\gamma}z'_{(2n+2m+1,2n+2m-2q),\gamma}\bar{y}_{(2n+2m+1,2n+1),\gamma}+\bar{y}_{(2m+2k+1,2k+1),\uparrow}x_{(2n,2n-2k),\uparrow} \nonumber
\eea
with $q=m,..,m+n-1$ and $k=q-m$.

\item 
$(2n)\circ(2m)\leftrightarrow (2n+2m-2q)\circ(2q)$, with 
$q=0,1..,m-1$:
\bea
&& \sum_{\gamma}\left[y_{(2n+2m+1,2n+2m-2q+1),\gamma}\bar{y}_{(2n+2m+1,2n+1),\gamma}+\bar{x}_{(2n+2m-2q,2n+1),\gamma}x_{(2m,2m-2q),\gamma}\right]\nonumber
\eea
\item $(2n)\circ(2m+1)_{\alpha}\leftrightarrow
(2n+2m-2q+1)_{\alpha}\circ(2q)$ with $q=0,1..,n+m$:
\bea
&&(1-\delta_{\alpha,\beta})x_{(2n+2m+2,2n+2m-2q),\beta}\bar{x}_{(2n+2m+2,2n+1),\beta}+\bar{y}_{(2n+2m-2q+1,2n+1),\alpha}y_{(2m+1,2m-2q+1),\alpha}\,\nonumber
\eea
for $q=0,..m$ and 
\bea
&&(1-\delta_{\alpha,\beta})\left[x_{(2n+2m+2,2n+2m-2q),\beta}\bar{x}_{(2n+2m+2,2n+1),\beta}+\bar{x}_{(2m+2k,2k+1),\beta}x_{(2n,2n-2k),\beta}\right]\,\nonumber
\eea
for $q=m+1,..,n+m$ and $k=q-m-1$.

\item
  $(2n)\circ(2m+1)_{\alpha}\leftrightarrow(2n+2m-2q)\circ(2q+1)_{\alpha}$
  with  $q=0,..,n+m-1$:
\bea
&&(1-\delta_{\alpha,\beta})\left[x_{(2n+2m+2,2n+2m-2q),\beta}\bar{x}_{(2n+2m+2,2n+1),\beta}+\bar{x}_{(2n+2m-2q,2n+1),\beta}t_{(2m+1,2m-2q),\beta}\right]
\nonumber\\
&&+\bar{x}_{(2n+2m-2q,2n+1),\beta}z_{(2m+1,2m-2q),\alpha}\,\nonumber
\eea
for $q=0,..m-1$ and 
\bea
&&(1-\delta_{\alpha,\beta})\left[x_{(2n+2m+2,2n+2m-2q),\beta}\bar{x}_{(2n+2m+2,2n+1),\beta}+\bar{t}_{(2m+2k+1,2k),\beta}x_{(2n,2n-2k+1),\beta}\right]
\nonumber\\
&&+\bar{z}_{(2m+2k+1,2k),\alpha}x_{(2n,2n-2k+1),\alpha}\,\nonumber
\eea
for $q=m+1,..,n+m$ and $k=q-m$.
\end{itemize}

\noindent \underline{potential (diagonal) terms}
\begin{itemize}  
\item 
$(2n)\leftrightarrow (2n)$:
\bea
&& \sum_{\alpha} \sum_{q=1}^{2n}
\bar{x}_{(2n,q),\alpha}x_{(2n,q),\alpha}+y_{(2n+1,1),\alpha}\bar{y}_{(2n+1,1),\alpha}+y_{(2n+1,2n+1),\alpha}\bar{y}_{(2n+1,2n+1),\alpha}\nonumber
\eea

\item  $(2n+1)_{\alpha}\leftrightarrow
(2n+1)_{\alpha}$:
\bea
&& \sum_{\alpha} \sum_{q=1,3,}^{2n+1}
\bar{y}_{(2n+1,q),\alpha}y_{(2n+1,q),\alpha}+\nonumber \\
&&+\sum_{\alpha} \sum_{q=2,4,}^{2n}\left[
   \bar{z}_{(2n+1,q),\alpha}z_{(2n+1,q),\alpha}
   +\bar{z}'_{(2n+1,q),\alpha}z'_{(2n+1,q),\alpha}
   +\bar{t}_{(2n+1,q),\alpha}t_{(2n+1,q),\alpha} \right]
\nonumber\\
&&+x_{(2n+2,1),\alpha}\bar{x}_{(2n+2,1),\alpha}+
   x_{(2n+2,2n+2),\alpha}\bar{x}_{(2n+2,2n+2),\alpha}
\nonumber.
\eea
\end{itemize}
\end{appendix}

\end{document}